\documentclass[aps,prl,twocolumn,nofootinbib,longbibliography,amsfonts,amssymb,amsmath,titlepage]{revtex4-1}

\usepackage[usenames,dvipsnames]{color}
\usepackage{microtype}
\usepackage{IEEEtrantools}
\usepackage{mathrsfs}
\usepackage{amsthm}
\usepackage{enumitem}
\usepackage{varwidth}
\usepackage[normalem]{ulem}
\usepackage{graphicx}
\usepackage{stackengine}
\usepackage{cprotect}
\usepackage[caption=false]{subfig}

\usepackage[pdftex]{hyperref}

\usepackage{tikz}
\usetikzlibrary{shapes.geometric, arrows.meta, backgrounds, fit, graphs, quotes}

\begin{document}

\title{Complete parameter inference for GW150914 using deep learning}

\author{Stephen R. Green}
\email{stephen.green@aei.mpg.de}
\author{Jonathan Gair}
\email{jonathan.gair@aei.mpg.de}
\affiliation{Max Planck Institute for Gravitational Physics (Albert Einstein Institute)\\
  Am M\"uhlenberg 1, 14476 Potsdam, Germany}

\begin{abstract}
  The LIGO and Virgo gravitational-wave observatories have detected
  many exciting events over the past five years. As the rate of
  detections grows with detector sensitivity, this poses a growing
  computational challenge for data analysis. With this in mind, in
  this work we apply deep learning techniques to perform fast
  likelihood-free Bayesian inference for gravitational waves. We train
  a neural-network conditional density estimator to model posterior
  probability distributions over the full 15-dimensional space of
  binary black hole system parameters, given detector strain data from
  multiple detectors. We use the method of normalizing
  flows---specifically, a \emph{neural spline} normalizing
  flow---which allows for rapid sampling and density estimation.
  Training the network is likelihood-free, requiring samples from
  the data generative process, but no likelihood evaluations. Through
  training, the network learns a \emph{global} set of posteriors: it
  can generate thousands of independent posterior samples per second
  for any strain data consistent with the prior and detector
  noise characteristics used for training.  By training with the
  detector noise power spectral density estimated at the time of
  GW150914, and conditioning on the event strain data, we use the
  neural network to generate accurate posterior samples consistent
  with analyses using conventional sampling techniques.
\end{abstract}

\maketitle

\emph{Introduction.---}Since the first detection in September
2015~\cite{Abbott:2016blz}, the LIGO/Virgo Collaboration has published
observations of gravitational waves from more than a dozen compact
binary
coalescences~\cite{LIGOScientific:2018mvr,LIGOScientific:2020stg,Abbott:2020uma,Abbott:2020khf},
primarily binary black hole mergers, but also two binary neutron star
mergers. In addition, the LIGO/Virgo Collaboration has publicly
released around fifty additional triggers~\cite{gracedb} of events of
interest, the details of which have so far not been published. These
observations have had a transformative impact on our understanding of
compact objects in the Universe, facilitated by inferring the
parameters of the system using accurate physical models of the emitted
gravitational waves.

This inference is extremely computationally expensive, as posterior
distributions of the parameters are usually obtained using more than
one waveform model to probe any potential systematic effects, and
using the physically most complete (and thus computationally most
costly) waveforms available. LIGO/Virgo currently employ Markov Chain
Monte Carlo and nested-sampling algorithms to obtain posterior
samples~\cite{Veitch:2014wba,Romero-Shaw:2020owr}. Run times for
single posterior calculations typically take days for binary black
hole systems and weeks for binary neutron
stars~\cite{LIGOScientific:2018mvr,Abbott:2018wiz}. These long run
times will become increasingly problematic as the sensitivity of the
instruments improves and event rates reach one per day or
higher~\cite{Aasi:2013wya}.

There is an urgent need for new approaches that can generate
scientific inferences much more rapidly than the existing
pipelines~\cite{Veitch:2014wba,Romero-Shaw:2020owr}. Deep-learning
methods are one promising approach to increase the speed of
gravitational wave inference by several orders of magnitude, that has
been receiving increasing focus in recent
years~\cite{Cuoco:2020ogp}. These techniques attempt to train a
neural-network conditional density estimator $q(\theta|s)$ to
approximate the Bayesian posterior distribution $p(\theta|s)$ of
parameter values $\theta$ given detector strain data $s$. Neural
networks typically have millions of parameters, which are optimized
stochastically during training to minimize an appropriate loss
function. With a ``likelihood-free'' training algorithm, it is never
necessary to draw samples from the posterior or evaluate a likelihood,
rather the procedure is generative and just requires an ability to
simulate data sets. Consequently, training can be done in a time
comparable to that taken to obtain posterior samples using a standard
method. There have been several previous works on this topic, but
these either simplified the description of the posterior, e.g., by
using a Gaussian approximation~\cite{Chua:2019wwt}, or simplified the
input, e.g., using a reduced space of parameters and a single
detector~\cite{Gabbard:2019rde}.

In a previous paper~\cite{Green:2020hst}, we used a type of neural
network known as a conditional variational autoencoder
(CVAE)~\cite{Kingma:2013,rezende2014stochastic} combined with
normalizing
flows~\cite{rezende2015variational,kingma2016improved,papamakarios2017masked,chen2016variational}
to learn the posterior distribution for inference of all parameters of
an aligned-spin quasi-circular merger observed with a single
gravitational-wave detector. With a single detector we could not
recover the full set of waveform parameters, and all data sets
analyzed were artificially generated with Advanced LIGO
design-sensitivity noise~\cite{zdhp}. In this Letter, we extend that
work into a tool that can, for the first time, be used to analyze real
data from the LIGO/Virgo interferometers. We describe a neural network
architecture, based on normalizing flows alone, that is able to
generate posteriors on the full $D=15$ dimensional parameter space of
quasi-circular binary inspirals, using input data from multiple
gravitational-wave detectors. We apply this network to analyze
observed interferometer data surrounding the first observed
gravitational-wave event, GW150914, and show that we can successfully
recover posterior distributions consistent with conventional
methods. This is the first demonstration that these methods can be
used in a realistic setting to produce fast and accurate scientific
inference on real data. This Letter thus establishes a new benchmark
in fast-and-accurate gravitational wave inference, as well as
describing methods that could also be applied to other inference
problems in experimental physics.

\emph{Neural network model.---}Our aim is to train a neural
conditional density estimator $q(\theta|s)$ to approximate the
gravitational-wave posterior $p(\theta|s)$. To this end, $q(\theta|s)$
must have sufficient flexibility to capture the detailed shape of the
true posterior over parameters $\theta$, as well as the dependence on
the complicated strain data $s$. We use the method of
\emph{normalizing flows}.

A normalizing flow $f$ is an invertible mapping on a sample space with
simple Jacobian determinant~\cite{rezende2015variational}. For a
conditional distribution, the flow must depend on $s$, so we denote it
$f_s$. The idea is to train the flow so that it maps a simple ``base''
distribution $\pi(u)$ into the far more complex $q(\theta|s)$. We
define the conditional distribution in terms of the flow by
\begin{equation}\label{eq:normalizing-flow}
  q(\theta|s) = \pi(f_s^{-1}(\theta)) \left| \det J_{f_s}^{-1} \right|,
\end{equation}
which is based on the change of variables rule for probability
distributions. The base distribution $\pi(u)$ should be chosen such
that it can be easily sampled and its density evaluated; we will
always take it to be standard multivariate normal of the same
dimension $D$ as the sample space.

By the properties of a normalizing flow, $q(\theta|s)$ inherits the
nice properties of the base distribution. Indeed, to draw a sample,
one first samples $u\sim\pi(u)$, and then sets $\theta = f_s(u)$; it
follows that $\theta \sim q(\theta|s)$. To evaluate the conditional
density, one uses~\eqref{eq:normalizing-flow}; the right hand side may
be evaluated by the defining properties that $f_s$ is invertible and
has simple Jacobian determinant.

\begin{figure}
  \begin{tikzpicture}
    [node distance=2cm,
     io/.style={rectangle, rounded corners, minimum width=.75cm, minimum height=.75cm,text centered, draw=black, fill=green!30},
     context/.style={ellipse, minimum width=1.5cm, minimum height=.75cm, text centered, draw=black, fill=blue!30},
     process/.style={rectangle, minimum width=2cm, minimum height=.75cm, text centered, draw=black, fill=red!30},
     point/.style={circle,inner sep=0pt,minimum size=2pt,fill=red},
     >={stealth},
     every new ->/.style={thick},
     flow/.style={thick, red}
     ]

     \matrix[row sep=5mm, column sep=5mm]{
       \node[io] (in) {$u$}; && \node[context, text width=2cm] (s)
       {strain $s$ \\ $n_{\text{SVD}}=100$}; \\
       
       \node[process,align=center] (p) {Permute,\\Linear}; & &\\

       \node[process] (split) {Coupling split}; &&\\
      
       \node[point] (uL){}; & & \node[process, align=left, text width=2.5cm] (nn)
       {Residual network\\
         \begin{itemize}[itemsep=0pt,partopsep=0pt,topsep=3pt,leftmargin=*]
         \item $n_{\text{hidden}}=512$
         \item $n_{\text{blocks}}=10$
         \item $K = 9$
         \end{itemize}
       };\\
       
       & \node[process, align=center] (spline) {Spline\\ transform}; &\\

       \node[process] (join) {Coupling join}; && \\
       
       \node[process,align=center]   (p2) {Permute,\\Linear}; &&\\
       
       \node[io]      (out)  {$\theta$}; &&\\
     };

     \graph [use existing nodes] {
       in <-> [flow] p <-> [flow] split <- [flow] uL -> [flow,edge label=$1:d$,swap] join <-> [flow] p2 <-> [flow] out;
       s -> nn;
       split <-> [flow, edge label=$d+1:D$, out=-75, in=105] spline <-> [flow, out=-125, in=55] join <-> [flow] p2;
       uL -> nn;
       nn -> [edge label={knots, derivatives}] spline;
     };
       
     \begin{scope}[on background layer]
       \node [draw=brown, fill=brown!10, fit=(p) (split) (nn) (uL) (spline) (join), label={[brown]-55:$\times n_{\text{flows}}=15$}] {};
     \end{scope}
   \end{tikzpicture}
   \caption{\label{fig:normalizing-flow}Overall structure of the
     normalizing flow~\cite{durkan2019neural} from the base space
     ($u$) to the parameter space ($\theta$), with optimal
     hyperparameter choices indicated. Red connections are
     invertible. The residual network is made up of
     $n_{\text{blocks}}$ residual blocks, each with two
     fully-connected hidden layers of $n_{\text{hidden}}$ units. Prior
     to each linear transformation~\cite{he2016identity}, we inserted
     batch normalization layers to speed
     training~\cite{ioffe2015batch} and Exponential Linear Units for
     nonlinearity~\cite{clevert2015fast}. Each block is also
     conditioned on the strain data $s$.}
\end{figure}
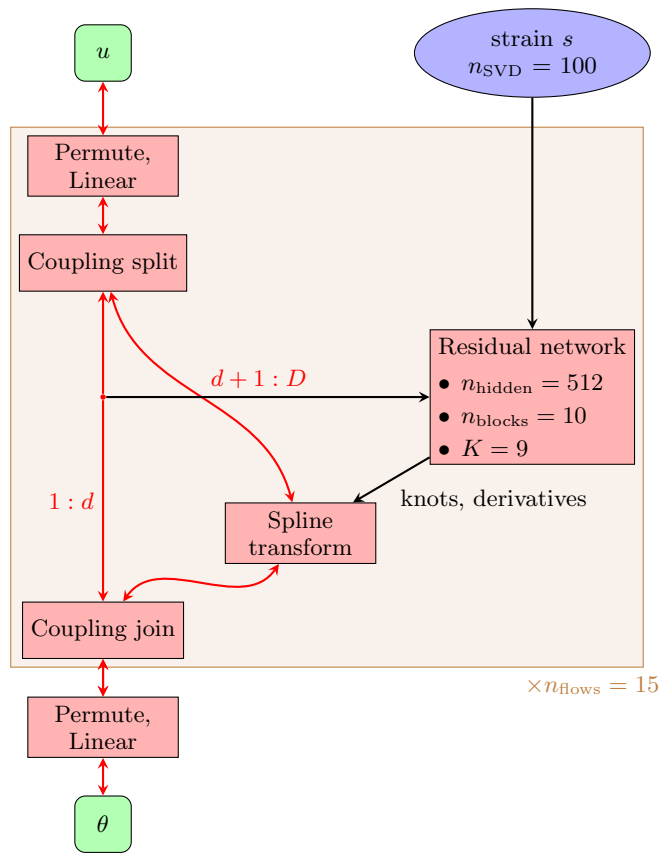

Normalizing flows are under active development in computer science,
and are usually represented by neural networks. Neural networks are
very flexible trainable function approximators, so they can give rise
to complex conditional densities. Our previous
work~\cite{Green:2020hst} used a Masked Autoregressive
Flow~\cite{papamakarios2017masked} with affine transformations. In the
present work, we use a much more powerful flow called a \emph{neural
  spline} flow~\cite{durkan2019neural}. We use directly the original
neural spline flow implementation~\cite{NSFgithub}, which we
illustrate in figure~\ref{fig:normalizing-flow}. We now give a brief
summary.

The flow is a composition of ``coupling transforms'' $c_s(u)$, each of
which transform elementwise half of the parameters (say, $u_{d+1:D}$)
conditional upon the other half ($u_{1:d}$) as well as the strain data
$s$~\cite{dinh2014nice}, i.e.,
\begin{equation}
  c_{s,i}(u) =
  \begin{cases}
    u_i & \text{if } i \le d,\\
    c_i(u_i; u_{1:d},s) & \text{if } i > d.
  \end{cases}
\end{equation}
If $c_i$ is invertible and differentiable with respect to $u_i$, then
it follows immediately that the coupling transform is a normalizing
flow. By composing $n_{\text{flows}}$ of these transforms, and
permuting the indices of $u$ in between, a much more flexible flow can
be obtained.

The neural spline coupling transform~\cite{durkan2019neural} takes
each $c_i$ to be a monotonically-increasing piecewise function,
defined by a set of knots $\{(u_i^{(k)},c_i^{(k)})\}_{k=0}^K$ and
positive-valued derivatives $\{\delta_i^{(k)}\}_{k=0}^K$, between
which are interpolated rational-quadratic (RQ) functions. The knots
and derivatives are output from a residual neural
network~\cite{he2015deep}, which takes as input $u_{1:d}$ and $s$;
details of the network are given in
figure~\ref{fig:normalizing-flow}. The RQ spline is differentiable and
has analytic inverse, so it satisfies the properties of a coupling
transform.

\emph{Training.---}The conditional density estimator $q(\theta|s)$
must be trained to approximate as closely as possible the
gravitational-wave posterior $p(\theta|s)$. We do this by tuning the
neural network parameters to minimize a loss function, the expected
value (over $s$) of the cross-entropy between the true and model
distributions,
\begin{align}\label{eq:L-cont}
  L &= - \int \mathrm{d}s\, p(s) \int \mathrm{d}\theta\, p(\theta|s) \log q(\theta|s) \nonumber\\
  &= - \int \mathrm{d}\theta\, p(\theta) \int \mathrm{d}s\, p(s|\theta) \log q(\theta|s).
\end{align}
On the second line we used Bayes' theorem to express $L$ in a form
that involves an integral over the likelihood rather than the
posterior; this is a key simplification which means posterior samples
are not needed for training. We evaluate the
integral~\eqref{eq:L-cont} on a minibatch of training data with a
Monte Carlo approximation,
\begin{equation}\label{eq:L-MC}
  L \approx - \frac{1}{N}\sum_{i=1}^N \log q(\theta^{(i)}|s^{(i)}),
\end{equation}
where $\theta^{(i)} \sim p(\theta)$, $s^{(i)}\sim p(s|\theta^{(i)})$,
and $N$ is the number of samples in the minibatch. We then use
backpropagation (the chain rule) to compute the gradient with respect
to network parameters, and minimize $L$ stochastically on minibatches
using the Adam optimizer~\cite{Kingma:2014vow}.

To obtain a training pair $(\theta^{(i)},s^{(i)})$, we draw
$\theta^{(i)}$ from the prior, we generate a waveform
$h(\theta^{(i)})$, and then we add a noise realization to obtain
$s^{(i)}$. Waveform generation is too costly to perform in real time
during training, so we adopt a hybrid approach: we sample
``intrinsic'' parameters in advance and save associated waveform
polarizations $h^{(i)}_{+,\times}$; at train time we sample
``extrinsic'' parameters, project onto detectors, and add noise. We
used $10^6$ sets of intrinsic parameters, which was sufficient to
avoid overfitting.

\emph{Prior.---}We perform inference over the full $D=15$ dimensional
set of precessing quasi-circular binary black hole parameters:
detector-frame masses $m_i$ $(i=1,2)$, reference phase $\phi_c$, time
of coalescence $t_{c,\,\text{geocent}}$, luminosity distance $d_L$,
spin magnitudes $a_i$, spin angles
$(\theta_i, \phi_{12}, \phi_{JL})$~\cite{Farr:2014qka}, inclination
angle $\theta_{JN}$, polarization angle $\psi$, and sky position
$(\alpha, \delta)$. Of these, we consider
$(m_i, \phi_c, a_i, \theta_i, \phi_{12}, \phi_{JL}, \theta_{JN})$ to
be intrinsic, so they are sampled in advance of training. To analyze
GW150914, we take a uniform prior over
\begin{subequations}
  \begin{IEEEeqnarray}{rCl}
  10~\mathrm{M_\odot} \le{}& m_i &{}\le 80~\mathrm{M_\odot},\\
  100~\mathrm{Mpc} \le{}& d_L &{}\le 1000~\mathrm{Mpc}, \\
  0 \le{}& a_i &{}\le 0.88,\\
  -0.1~\mathrm{s} \le{}& t_{c,~\text{geocent}} &{}\le 0.1~\mathrm{s}.
  \end{IEEEeqnarray}
\end{subequations}
The prior is standard over the remaining quantities. We take
$t_{c,~\text{geocent}}=0$ to be the trigger time, and we require
$m_1\ge m_2$.

Although a prior uniform in the comoving source
frame~\cite{Romero-Shaw:2020owr} would be most physical, we adopted
a uniform prior over $d_L$ and an upper bound of 1000~Mpc to more
uniformly cover the parameter space and improve training. We applied
the physical prior in postprocessing by reweighting samples. Also to
improve training, we rescaled all parameters to have zero mean and
unit variance.

\emph{Strain data.---}For likelihood-free training, we require
simulated strain data sets $s^{(i)}$ that arise from the data
generative process, $s^{(i)} \sim p(s|\theta^{(i)})$. We assume
stationary Gaussian noise, so the gravitational-wave likelihood
function is known explicitly, but the likelihood-free approach applies
even when this is not the case---e.g., in the presence of non-Gaussian
noise---as long as it is possible to simulate data.

We generate training waveforms using the \verb|IMRPhenomPv2|
frequency-domain precessing
model~\cite{Hannam:2013oca,Khan:2015jqa,Bohe:2016}. We take a
frequency range of [20, 1024]~Hz, and a waveform duration of 8~s. We
then whiten $h^{(i)}_{+,\times}$ using the noise PSD estimated from
1024~s of detector data prior to the
event. Following~\cite{Chua:2019wwt}, we compress the whitened
waveforms to a reduced-order representation; we use a singular value
decomposition (SVD), and keep the first $n_{\text{SVD}}=100$ components.

At train time, we sample extrinsic parameters and generate detector
signals. This requires a trivial rescaling to apply $d_L$, and linear
combinations of $h^{(i)}_{+,\times}$ to project onto the antenna
patterns for the two LIGO detectors. To apply time delays in the
reduced-basis (RB) representation, we follow the approach
of~\cite{Canizares:2013ywa,Smith:2016qas} of pre-preparing a grid of
time-translation matrix operators that act on vectors of RB
coefficients, using cubic interpolation for intermediate times. Since
the transformation to RB space is a rotation, we add white noise
directly to the RB coefficients of the whitened waveforms to obtain
$s^{(i)}$. Finally, data is standardized to have zero mean and unit
variance in each component.

\emph{Results.---}We trained for 500 epochs with a batch size of
512. The initial learning rate was 0.0002, and we used cosine
annealing~\cite{loshchilov2016sgdr} to reduce the learning rate to
zero over the course of training. We performed a search over network
hyperparameters, and have listed those with best performance (as
measured by final validation loss) in
figure~\ref{fig:normalizing-flow}. During training, we reserved $10\%$
of our training set for validation, and found no evidence of
overfitting. With an NVIDIA Quadro P4000 GPU, training took $\approx6$
days.

\begin{figure*}
  \stackinset{r}{}{t}{}{\includegraphics[width=0.45\textwidth]{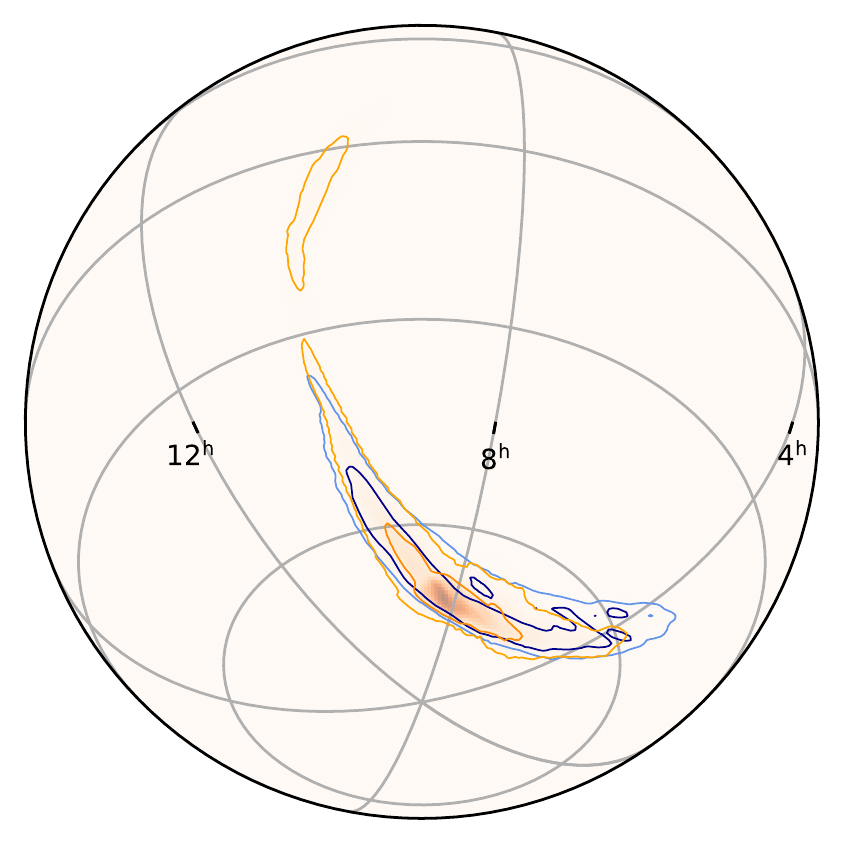}}{
    \includegraphics[width=\textwidth]{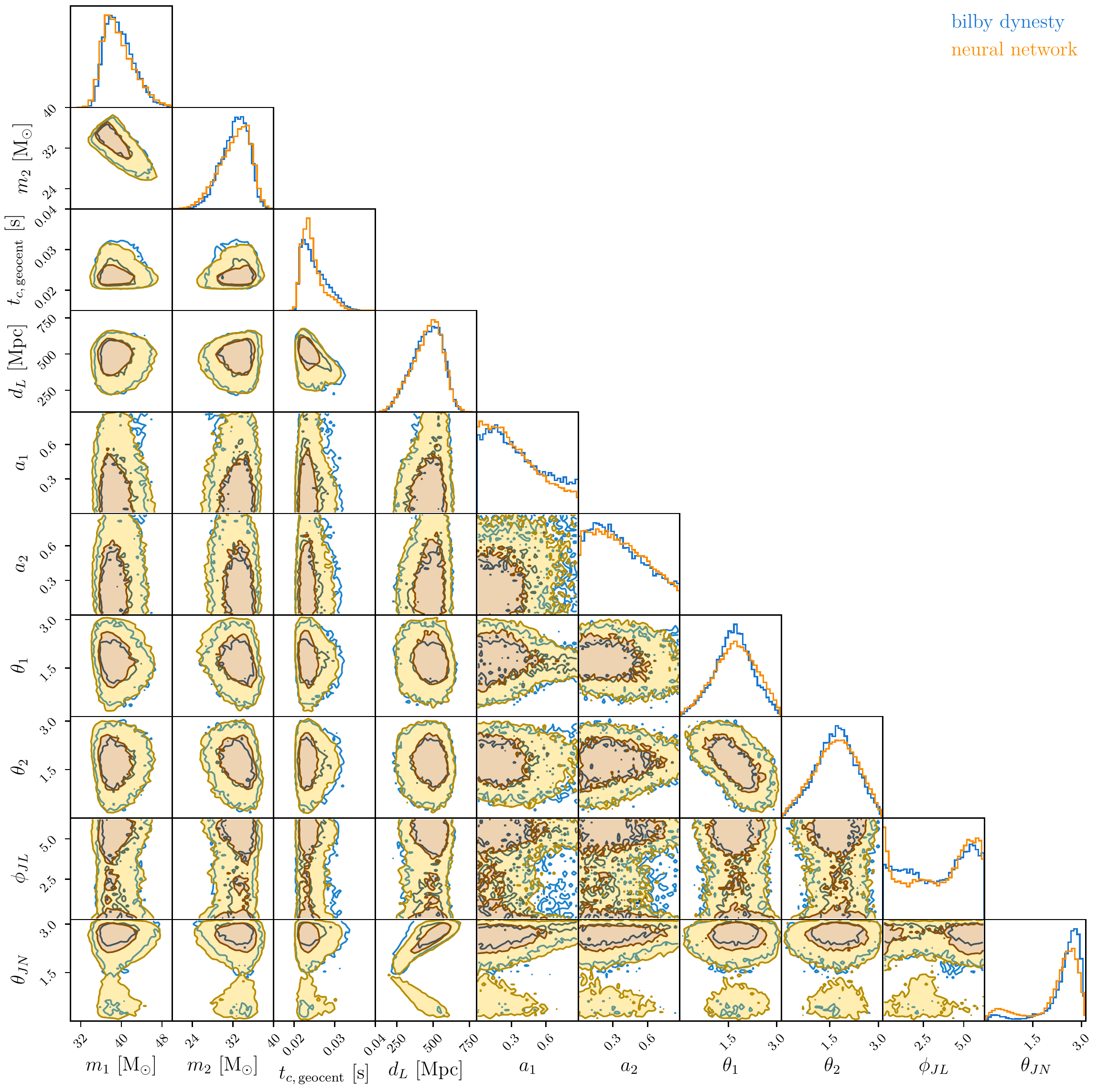}}
  \cprotect\caption{\label{fig:posterior}Marginalized one- and two-
    dimensional posterior distributions over a subset of parameters,
    comparing the normalizing flow (orange) and \verb|bilby|
    \verb|dynesty| (blue). Contours represent 50\% and 90\% credible
    regions. Neural network posteriors are constructed from
    $5\times10^4$ samples. The inset shows the sky position, with
    rejection sampling used to obtain unweighted neural network
    samples.}
\end{figure*}

To perform inference on GW150914, we took 8~s of detector data
containing the signal and expressed it in the RB representation. We
then drew samples from the base space, and applied the normalizing
flow conditioned on the strain data to obtain samples from
$q(\theta|s)$. This produced samples at a rate of 5,000 per second. We
benchmarked these against samples produced by
\verb|bilby|~\cite{Ashton:2018jfp,Romero-Shaw:2020owr} with the
\verb$dynesty$ sampler~\cite{Speagle_2020}.

Our main result is presented in figure~\ref{fig:posterior}, which
compares the neural network and \verb|bilby| posteriors. Both
distributions are clearly in very close agreement. There are minor
differences in the inclination angle $\theta_{JN}$, where the neural
network gives more support to the secondary mode, and the sky
position, where the neural network also develops a small secondary
mode. With more training or a larger network, we expect even better
convergence.

\begin{figure}
  \includegraphics[width=\linewidth]{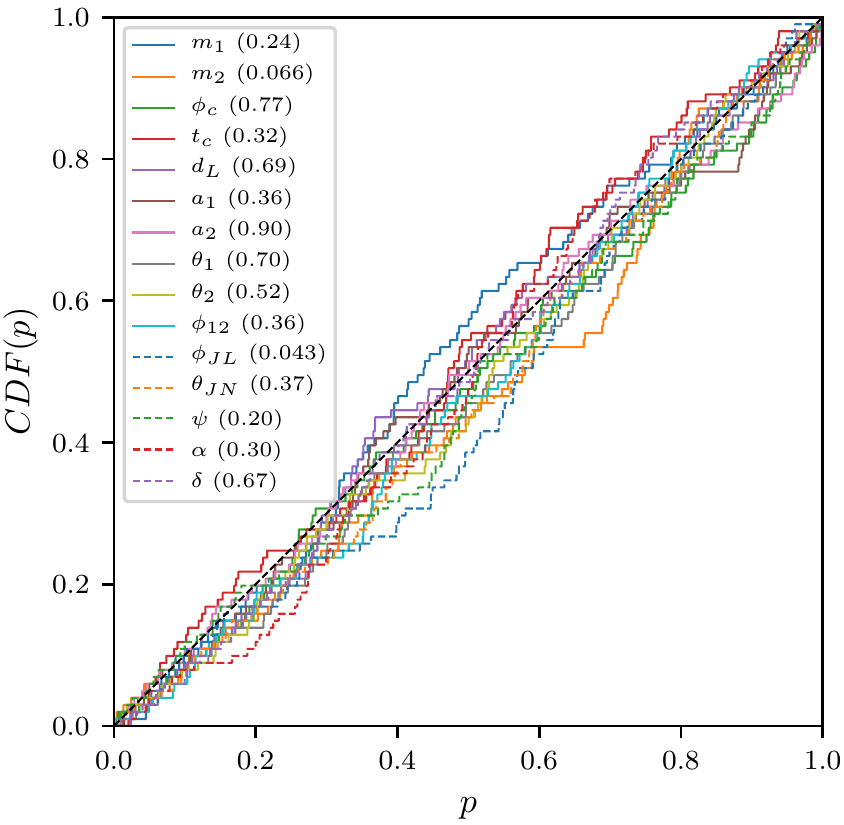}
  \caption{\label{fig:pp}P--P plot for 100 artificial strain data sets
    analyzed by the neural network. For each injection and
    one-dimensional posterior distribution, we compute the percentile
    value of the injected parameter. The figure shows the CDF of the
    injections for each parameter, which should lie close to the
    diagonal if the network is performing properly. KS test $p$-values
    are given in the legend.}
\end{figure}

Although our demonstration has focused on GW150914, the neural network
has been trained to generate any posterior consistent with the prior
and the given noise PSD. To demonstrate this, performed inference on
100 injections of waveforms drawn from the prior with added
noise. Performance is summarized in the P--P plot shown in
figure~\ref{fig:pp}. This shows the cumulative distribution function
(CDF) of the percentile scores of each injection parameter within the
one-dimensional marginalized posteriors.  The percentiles should be
distributed uniformly between 0 and 1; since the CDF lies close to the
diagonal, we conclude that the network is properly sampling the
posteriors. This is confirmed by a Kolmogorov-Smirnov (KS) test.

In our experiments, we also varied $n_{\text{SVD}}$, and we found
slightly reduced performance as this was increased. This indicates
that, although with less compression it should be possible to produce
tighter posteriors, better network optimization is required to take
full advantage. Indeed, subleading SVD elements contain mostly noise,
which makes training more difficult.

\emph{Conclusions.---}In this Letter, we have demonstrated for the
first time that deep neural networks can accurately infer all 15
binary black hole parameters from real gravitational-wave strain
data. Once the network is trained, inference is extremely fast,
producing 5,000 independent samples per second.

Rapid parameter estimation is critical for multimessenger followup and
for confronting the expected high rate of future detections.  An
advantage of likelihood-free methods is that waveform generation is
done in advance of training and inference, rather than at sampling
time as for conventional methods. Thus, waveform models that include
more physics but may be slower to evaluate~\cite{Ossokine:2020kjp} can
be used to analyze data in the same time as faster models.

The network we presented is tuned to a particular noise PSD---in this
case, estimated just prior to GW150914. However, the noise
characteristics of the LIGO and Virgo detectors vary from event to
event, and ultimately we would like amortize training costs by
building a conditional density estimator that can do inference on any
event without retraining for each PSD. One approach would be to
condition the model on PSD information: during training, waveforms
would be whitened with respect to a PSD drawn from a distribution
representing the variation in detector noise from event to event, and
(a summary of) this PSD information would be passed to the network as
additional context. (PSD samples can be obtained from detector data at
random times.) For inference, PSD information would then be passed
along with the whitened strain data. Similar approaches could also be
used to treat non-Gaussian noise artifacts.

In contrast to CVAEs used in past
work~\cite{Gabbard:2019rde,Green:2020hst}, normalizing flows have the
advantage of estimating the density directly, without any need to
marginalize over latent variables. This means that the loss function
can be taken to be the cross-entropy~\eqref{eq:L-MC} rather than an
upper bound~\cite{Kingma:2013,rezende2014stochastic}. Moreover, since
$q(\theta|s)$ is a normalized probability distribution, the Bayesian
evidence can be obtained as a byproduct. The performance we achieved
without latent variables in this work was made possible by the use of
a more powerful normalizing flow~\cite{durkan2019neural} compared
to~\cite{Green:2020hst}. As new and more powerful normalizing flows
are developed by computer scientists in the future, they will be
straightforward to deploy to further improve the performance and
capabilities of deep learning for gravitational-wave parameter
estimation.

\medskip

We would like to thank C. Simpson for helpful discussions and for
performing training runs during the early stages of this work. Our
code was implemented in \verb|PyTorch|~\cite{NEURIPS2019_9015}, and
with the neural spline flow implementation of~\cite{NSFgithub}. Plots
were produced with \verb|matplotlib|~\cite{Hunter:2007},
\verb|ChainConsumer|~\cite{Hinton2016} and
\verb|ligo.skymap|~\cite{Singer:ligoskymap}.

\bibliography{mybib.bib}

\end{document}